\documentclass[10pt]{article}
\RequirePackage{titlesec}
\RequirePackage{amsmath,esdiff}
\RequirePackage{amssymb}
\RequirePackage{authblk}

\RequirePackage[normalem]{ulem}
\RequirePackage{bbold,bm}
\RequirePackage{BOONDOX-cal}
\RequirePackage[colorlinks=true,linkcolor=blue,citecolor=red,urlcolor=red]{hyperref}
\RequirePackage{geometry}
\RequirePackage{setspace}
\RequirePackage{cite}
\RequirePackage{flushend}
\RequirePackage{mathrsfs}
\RequirePackage{hyperref}
\RequirePackage{graphicx}
\RequirePackage{cancel}
\RequirePackage{array,esint}
\RequirePackage[most]{tcolorbox}

\graphicspath{}
\titleformat{\section}{\fontsize{12}{12}\bfseries}{\thesection}{1em}{}
\begin{document}
\title{\bf Lewis and Berry phases for a gravitational wave interacting with a quantum harmonic oscillator}
\vskip 1cm
	\vskip 1cm
\author{\textbf{$\mathbf{Soham}$ $\mathbf{Sen}^{\dagger}$, $\mathbf{Manjari}$ $\mathbf{Dutta}^{*}$ and $\mathbf{Sunandan}$ $\mathbf{Gangopadhyay}^{\ddagger}$
\footnote{{\quad}\\
{${}^\dagger$sensohomhary@gmail.com, soham.sen@bose.res.in}\\
{${}^*$chandromouli15@gmail.com}\\
{${}^\ddagger$sunandan.gangopadhyay@bose.res.in, sunandan.gangopadhyay@gmail.com}}}}
\affil{{Department of Astrophysics and High Energy Physics}\\
{S.N. Bose National Centre for Basic Sciences}\\
{JD Block, Sector III, Salt Lake, Kolkata 700

106, India}}
\date{}
\maketitle
\begin{abstract}
\noindent In this work, we compute the Lewis and Berry phases for a gravitational wave interacting with a two dimensional quantum harmonic oscillator in the transverse-traceless gauge. We have considered a gravitational wave consisting of the plus polarization term only. Considering the cross polarization term to be absent makes the Hamiltonian separable in terms of the first and the second spatial coordinates. We then compute the Lewis phase by assuming a suitable form of the Lewis invariant considering only quadratic order contributions from both position and momentum variables. 
Next, we obtain two Lewis invariants corresponding to each separable part of the full Hamiltonian of the system. Using both Lewis invariants, one can obtain two Ermakov-Pinney equations, from which we finally obtain the corresponding Lewis phase. Then making an adiabatic approximation enables us to isolate the Berry phase for the full system. After this we obtain some explicit expressions of the Berry phase for a plane polarized gravitational wave with different choices of the harmonic oscillator frequency.   Finally, we consider a gravitational wave with cross polarization only interacting with an isotropic two dimensional harmonic oscillator. For this we obtain the Lewis phase and the total Berry phase of the system, which is found to be dependent upon the cross polarization part of the gravitational wave. 
\end{abstract}
\section{Introduction}
In 1916, Albert Einstein predicted that gravitational radiation must exist by linearizing the general theory of relativity. This was the birth of gravitational waves. The detection of gravitational wave in 2015 (from two colliding neutron stars), led to an upsurge in research related to gravitational waves and its various aspects almost after a century had passed from the initial theoretical prediction in 1916. From its initial detection, several theoretical models have been proposed to investigate the effects of gravitational waves in several quantum mechanical scenarios. Although it is theoretically more prudent to consider the background geometry to be curved as most of the gravitational wave detectors (LIGO and VIRGO) are ground based detectors, the usual theoretical notion is to consider the gravitational wave as a fluctuation over a flat Minkowski background. 
The idea of observing gravitational waves directly was pioneered by Weber in the 1960s. He for the first time introduced the so-called resonant bar detectors \cite{Weber, Weber2}, and claimed that gravitational waves have been detected. Thereafter, there has been a monumental effort to increase the sensitivity of these detectors. The possibility of increasing their sensitivity to higher levels has also led to the idea of probing quantum gravity effects in these setups \cite{SG01,SG02,SG03,GBar1,GBar2,GBar3,OTM,ResonantOTM}. The basic model in the resonant bar detector setup can be interpreted as a quantum mechanical forced gravitational wave harmonic oscillator system. The incoming gravitational wave interacts with the elastic matter. The tiny vibrations in the detector are termed phonons.


The analysis involving the interaction between a gravitational wave and a quantum harmonic oscillator is quite well known in the recent literature. The interaction of classical gravitational waves with matter was studied in \cite{Speliotopoulos} using a quantum mechanical approach. The study of time dependent quantum harmonic oscillators is another area of research that is of immense interest among the scientific community as well. This started with the work of Lewis \textit{et. al.} \cite{LewisRiesenfeld}.  The method involves using exact invariants to critically study these systems \cite{Lewis2,Lewis3}. There have been numerous analyses involving the damping in a one dimensional quantum harmonic oscillator \cite{Pedrosa1,Jannussis,Pedrosa2,Abdalla,Pedrosa3}. There also have been few studies involving damping in a two dimensional quantum harmonic oscillator \cite{Lawson} and later extended to noncommutative space \cite{Manjari}. An explicit analysis of a damped harmonic oscillator with time dependent frequency in a noncommutative space in the presence of magnetic field was done in \cite{Manjari2}.

On the other hand in a process if a very gradual change of external parameters occurs \cite{BornFock,Griffiths}, then we call it an adiabatic process with which two characteristic time scales are involved. These two characteristic time scales are termed as ``\textit{external time}" and ``\textit{internal time}". The ``\textit{external time}" signifies the parameters (of the system), over which the system changes significantly and the ``\textit{internal time}" is related to the motion of the system. For an adiabatic process, the ``\textit{external time}" is dominant over the internal time which signifies that the Hamiltonian parameters evolve quite slowly over time. Computing the Lewis phase and doing the adiabatic approximation, gives an efficient way to obtain this phase. 

In \cite{Berry1,Berry2}, Berry used a Hamiltonian with more than one time-dependent parameter. One of the criteria he used in his work was that the eigenfunctions of the Hamiltonian must be complex numbers. Now if the set of parameters changes adiabatically for such a system through a closed path (such that it can return to its initial position), then the system acquires a geometric phase \cite{Anandan} and for these systems, the time reversal symmetry of the Hamiltonian must be broken \cite{Ghosh}. He showed that along with the time dependent dynamical phase, a time independent and path dependent geometric phase is also picked up by the system. This path dependent geometric phase is also known as the ``\textit{Berry phase}" which is a physically measurable quantity. As it is not as large as compared to the dynamical phase, it is possible to pick up the geometric phase using sophisticated experiments \cite{Berryexp1,Berryexp2}. The connection between Berry’s phase and Lewis’s phase for the quadratic Hamiltonians was first explored in the work \cite{Morales}. Few more interesting investigations of such phases have been performed especially in the case of quantum harmonic oscillators \cite{Fring,Giavarini1,Giavarini2,Giavarini3,Dittrich,Manjari3,Bibhas}. Information theoretic viewpoint of the computation of such phases can be found in \cite{Jimenez,Zanardi} .


In this study, our primary objective is to obtain Berry's geometric phase for the non-relativistic quantum harmonic oscillator interacting with gravitational waves. The motivation for looking at Berry phases in this kind of setup is that they can prove to be a very strong detection tool for gravitational waves. Our approach to looking for the Berry phase employs the Lewis method \cite{LewisRiesenfeld}. This technique offers an elegant way to derive the Berry phase by applying the adiabatic approximation once the Lewis phase has been obtained. Here,  we consider a gravitational wave in the transverse-traceless gauge interacting with a two dimensional time dependent quantum harmonic oscillator with time dependent frequency. As a result, the system Hamiltonian becomes time dependent. We have then followed the analysis by Lewis \textit{et. al.} \cite{LewisRiesenfeld} to obtain an invariant corresponding to the total Hamiltonian of the system.   The analysis provides a direct connection between the eigenstates of the time-dependent invariant and the solution of the Schr\"{o}dinger's equation. This Lewis invariant method is very elegant in the sense that it needs no prior approximation to obtain the Lewis phase corresponding to the system and the geometric part of the phase is obtained using a standard adiabatic approximation which is the Berry phase of the system \cite{Morales}. Apart from this Lewis invariant approach one can resort to a path integral approach for obtaining Berry's geometric phase as well \cite{Hammann,Sakurai}. We start our analysis by considering the plus polarization of the gravitational wave. We find out that the Hamiltonian can be decoupled into two individual Hamiltonians corresponding to the two spatial directions. We then obtain the Lewis phase and eventually the Berry phase of the two-dimensional harmonic oscillator-gravitational wave system. The importance of looking for the Berry phase in the quantum harmonic oscillator-gravitational wave system is evident. Finding a non-trivial detectable Berry phase can prove to be important in the detection of gravitational waves in resonant bar detector systems \cite{GBar1,GBar2,GBar3,OTM}.  The most important insight that we obtain via this analysis is that the Lewis phase, which can be considered to be the total phase of the system, has no contribution from the gravitational wave in its dynamic part after applying the adiabatic approximation. The contribution of the gravitational wave gets captured entirely in the geometric part of the Lewis phase. After that we have considered a gravitational wave, carrying cross polarization only, interacting with a two dimensional and isotropic harmonic oscillator system. We observe that the Hamiltonian can be decoupled as the plus polarization case by considering a simple change of coordinates which effectively rotates the coordinate system by an angle $\frac{\pi}{4}$ in the clockwise direction. Again by following the same analysis, we obtain the Lewis phase and eventually the total Berry phase of the system. Finally, in an appendix, we have tried to obtain the Lewis invariant for a two dimensional harmonic oscillator system interacting with a gravitational wave carrying both plus and cross polarizations. Due to the presence of some self-contradictory constraint equations, we were unable to obtain a valid form of the Lewis invariant for the same. Hence, we keep it as an open problem and wish to follow up in future work.  


The organization of the paper goes as follows. In section \ref{S2}, we discuss the Hamiltonian of the system. In section \ref{S3}, we have discussed in brief the method of Lewis invariant and obtained the Lewis invariant corresponding to our system in section \ref{S4}. In section \ref{S5}, we have obtained the creation and annihilation operators from the Lewis invariant obtained in section \ref{S4}. In section \ref{S6}, we have obtained the Lewis and the Berry phases for our system Hamiltonian. Next in section \ref{S7}, we have obtained a few explicit Berry phases for some distinct set of choices of the Harmonic oscillator frequencies. In section \ref{S7b}, we have calculated the contribution of the boundary term towards the total geometric phase of the system. In section \ref{S8a}, we have analyzed the case of a gravitational wave with only a cross polarization interacting with an isotropic two dimensional harmonic oscillator and obtained the corresponding Lewis and Berry phases for the system. Finally, we have concluded the paper in section \ref{S8}.
\section{Hamiltonian of the system}\label{S2}
\noindent In this paper we consider a two dimensional simple harmonic oscillator interacting with a gravitational wave in the transverse-traceless gauge. The background metric in the linearized approximation can then be expressed as the flat Minkowski metric plus the perturbation due to the gravitational wave
\begin{equation}\label{2.1}
g_{\mu\nu}=\eta_{\mu\nu}+h_{\mu\nu}
\end{equation}
where $\eta_{\mu\nu}=\text{diag}\{-1,1,1,1\}$ with $\mu,\nu=\{0,1,2,3\}$. The form of $h_{\mu\nu}$ in the transverse traceless gauge is given in the matrix form as 
\begin{equation}\label{2.2}
h_{\mu\nu}=\begin{pmatrix}
0&0&0&0\\
0&h_+(t)&h_\times(t)&0\\
0&h_\times(t)&-h_+(t)&0\\
0&0&0&0
\end{pmatrix}
\end{equation} 
where $h_+(t)=2\chi(t)\varepsilon_+$ and $h_\times(t)=2\chi(t)\varepsilon_\times$, with $\varepsilon_+$ and $\varepsilon_\times$ denoting the plus and the cross polarization of the gravitational waves having amplitude $\chi(t)$. If we now define the polarization tensor in three spatial dimensions to be $\varepsilon_l=(\varepsilon_\times,0,\varepsilon_+)$, we can write down the fluctuation term in terms of the Pauli spin matrix elements $\sigma^{l}_{jk}$ in the following way
\begin{equation}\label{2.3}
\begin{split}
h_{jk}(t)&=2\chi(t)\left(\varepsilon_+\sigma^3_{jk}+\varepsilon_\times\sigma^1_{jk}\right)\\
&=2\chi(t)\varepsilon_l\sigma^{l}_{jk}
\end{split}
\end{equation}
where $l=\{1,2,3\}$ and $j,k=\{1,2\}$. To write down eq.(\ref{2.3}) from $h_{\mu\nu}$, it is important to note that the matrix $h_{\mu\nu}$ can be reduced effectively to $2\times 2$ matrix spanning the $x$ and $y$ directions only. Before explicitly making use of the gravitational wave perturbation term in eq.(\ref{2.3}), we shall write down the classical form of harmonic oscillator-gravitational wave interaction Hamiltonian (with time dependent frequency) as \cite{SG01}
\begin{equation}\label{2.8}
H=\frac{1}{2m}p_kp^k+\Gamma^{j}_{~0k}x^kp_j+\frac{1}{2}m\omega^2(t)x_kx^k
\end{equation}
where up to $\mathcal{O}(h)$ we can express $\Gamma^{j}_{~0k}\cong\frac{1}{2}\eta^{jm}\partial_0h_{mk}$ ($j,k,m\simeq1,2$). In order to obtain the above Hamiltonian of the system, one needs to start from the action given by
\begin{equation}\label{2.4}
S=\int dt\left(\frac{1}{2}m\dot{x}_k\dot{x}^k-\frac{1}{2}mR^{j}_{~0k0}x_jx^k-V\right)
\end{equation}
 where $j,k=\{1,2\}$ and $V$ is the harmonic oscillator potential which is given by
\begin{equation}\label{2.5}
V=\frac{1}{2}m\omega^2x_kx^k~.
\end{equation}
In general, it is more convenient to consider $\omega$ as a constant but in our analysis we consider the frequency of the harmonic oscillator to be time dependent ($\omega=\omega(t)$). Dropping the boundary term, we can recast the form of the action in eq.(\ref{2.4}) given as
\begin{equation}\label{2.6}
S=\int dt\left(\frac{1}{2}m\dot{x}_k\dot{x}^k-m\Gamma^{j}_{~0k}\dot{x}_jx^k-V\right)
\end{equation}
where $R^{j}_{~0k0}=-\partial_0\Gamma^{j}_{~0k}$ and the modified Lagrangian of the system is given by
\begin{equation}\label{2.7}
L'=\frac{1}{2}m\dot{x}_k\dot{x}^k-m\Gamma^{j}_{~0k}\dot{x}_jx^k-V~.
\end{equation}
From this modified Lagrangian, one gets back our starting Hamiltonian in eq.(\ref{2.8}). Before proceeding further, we make a comment about the boundary term that we have dropped in eq.(\ref{2.6}) to obtain the modified form of the action in eq.(\ref{2.6}). The boundary term has the form given as
\begin{equation}\label{2.9a}
\mathcal{B}=\frac{m}{2}\int_0^t d\tau\frac{d}{d \tau}\left(
\Gamma^j_{~0k}(\tau)x_j(\tau)x^k(\tau)\right)~.
\end{equation}
Note that the boundary term may contribute towards the overall geometric phase. We shall calculate this contribution in section \ref{S7b}.
Using eq.(\ref{2.3}) in eq.(\ref{2.8}) and raising the phase space variables to operator status, we can finally write down the Hamiltonian for the quantum harmonic oscillator to be
\begin{equation}\label{2.9} 
\begin{split}
\hat{H}(t)=&\frac{a(t)}{2}(\hat{p}_1^2+\hat{p}_2^2)+\frac{b(t)}{2}(\hat{x}_1^2+\hat{x}_2^2)+d(t)(\hat{x}_1\hat{p}_1+\hat{p}_1\hat{x}_1)-d(t)(\hat{x}_2\hat{p}_2+\hat{p}_2\hat{x}_2)+f(t)(\hat{x}_1\hat{p}_2+\hat{p}_1\hat{x}_2)
\end{split}
\end{equation}
where $a(t)=\frac{1}{m}$, $b(t)=m\omega^2(t)$, $d(t)=\frac{1}{2}\varepsilon_+\dot{\chi}(t) $, and $f(t)=\varepsilon_\times \dot{\chi}(t)$. It is important to note that $d(t),f(t)\sim \mathcal{O}(h)$. 
Instead of progressing further with the form of the Hamiltonian $H(t)$ in eq.(\ref{2.9}), we simplify the form of the above Hamiltonian by setting $f(t)=0$ in eq.(\ref{2.9}). This can be done without any loss of generality, since removing the $f(t)$ term, implies looking at the gravitational wave interacting with the two dimensional harmonic oscillator having plus polarization only. Later in section (\ref{S8a}), we have considered the case where the  gravitational wave can be considered to have the cross polarization only without any loss of generality\footnote{We have discussed the case where  the gravitational wave carries both polarizations, in the Appendix.}. 
Apart from setting $f(t)=0$ (which is equivalent to setting $\varepsilon_\times=0$), any alternative choice, such as $f(t) = $   constant, would result in setting the factor $\dot{\chi}(t)$ to a constant. This in turn would imply  $d(t)$=constant. As we shall see in section \ref{S6},  the Berry phase of the system depends explicitly upon the factor $\dot{d}(t)$, which therefore becomes zero when $f(t)$=constant.

\noindent Setting $f(t)=0$, the Hamiltonian in eq.(\ref{2.9}) can now be recast in the following form, 
 \begin{equation}\label{3.10}
\begin{split}
\hat{H}_{+}(t)=&\frac{a}{2}\hat{p}_1^2+\frac{b(t)}{2}\hat{x}_1^2+d(t)(\hat{x}_1\hat{p}_1+\hat{p}_1\hat{x}_1)+\frac{a}{2}\hat{p}_2^2+\frac{b(t)}{2}\hat{x}_2^2-d(t)(\hat{x}_2\hat{p}_2+\hat{p}_2\hat{x}_2)\\
=&\hat{H}_1(t)+\hat{H}_2(t)
\end{split}
\end{equation}
where 
\begin{align}
\hat{H}_1(t)&=\frac{a}{2}\hat{p}_1^2+\frac{b(t)}{2}\hat{x}_1^2+d(t)(\hat{x}_1\hat{p}_1+\hat{p}_1\hat{x}_1)~,\label{3.11a}\\
\hat{H}_2(t)&=\frac{a}{2}\hat{p}_2^2+\frac{b(t)}{2}\hat{x}_2^2-d(t)(\hat{x}_2\hat{p}_2+\hat{p}_2\hat{x}_2)~.\label{3.11b}
\end{align}

\noindent Note that, we can effectively write down the Hamiltonian as a sum of two Hamiltonians describing the dynamics of the system corresponding to each spatial dimension. We will now try to find out the Lewis invariant corresponding to the Hamiltonian given in eq.(\ref{3.10}). 
\section{Basic introduction to Lewis invariant}\label{S3}
\noindent The model Hamiltonian obtained in eq.(\ref{2.9}) is an explicit function of time. Following the analysis in \cite{LewisRiesenfeld}, it may be possible to construct a time dependent Hermitian operator $\hat{I}(t)$ such that
\begin{equation}\label{3.1}
\begin{split}
\frac{d\hat{I}}{dt}=\frac{\partial \hat{I}}{\partial t}+\frac{1}{i\hbar}[\hat{I},\hat{H}]=0~.
\end{split}
\end{equation}
If a time dependent Schr\"{o}dinger state vector $|\psi\rangle$ satisfies the following relation
\begin{equation}\label{3.2}
i\hbar\frac{\partial|\psi_k\rangle}{\partial t}=\hat{H}|\psi_k\rangle
\end{equation}
then we can deduce the following relation (by making use of eq.(\ref{3.1}))
\begin{equation}\label{3.3}
\begin{split}
i\hbar\frac{\partial}{\partial t}(\hat{I}|\psi_k\rangle)=&i\hbar\frac{\partial \hat{I}}{\partial t}|\psi_k\rangle+\hat{I}\left[i\hbar\frac{\partial |\psi_k\rangle}{\partial t}\right]\\=&\left[i\hbar\frac{\partial \hat{I}}{\partial t}+\hat{I}\hat{H}\right]|\psi_k\rangle\\
=&\hat{H}(\hat{I}|\psi_k\rangle)
\end{split}
\end{equation}
where to obtain the last line of the above equation, we have made use of eq.(\ref{3.1}). Eq.(\ref{3.3}) suggests that the action of $\hat{I}(t)$ on the Schr\"{o}dinger state vector creates another state vector. If one now assumes that the invariant operator is one among a complete set of commuting observables, then it is straightforward to find out a complete set of eigenstates for the Hermitian invariant operator $\hat{I}(t)$. We assume that for the invariant in our case, the eigenstate is $|\varphi_k\rangle$.  Then we can write down the following relation
\begin{equation}\label{3.4}
\hat{I}(t)|\varphi_k\rangle=\zeta|\varphi_k\rangle
\end{equation}
with $\zeta$ being the eigenvalue. Following the analysis in \cite{LewisRiesenfeld}, one can write down the time dependent solution of the Schr\"{o}dinger equation in eq.(\ref{3.2}) in terms of the eigenstate $|\varphi_k\rangle$ as follows
\begin{equation}\label{3.5}
|\psi_k\rangle=e^{i\theta_k(t)}|\varphi_k\rangle
\end{equation} 
where $\theta_k(t)$ is a real function of time. Substituting eq.(\ref{3.5}) back in the time dependent Schr\"{o}dinger equation given in eq.(\ref{3.2}), we obtain the following relation
\begin{equation}\label{3.5a}
e^{i\theta_k(t)}\left(i\hbar\partial_t-\hbar\dot{\theta}_k(t)\right)|\phi_k\rangle=e^{i\theta_k(t)}\hat{H}|\phi_k\rangle~.
\end{equation}
We can recast eq.(\ref{3.5a}) via the action of $\langle \psi_k|$ which gives
\begin{equation}\label{3.6a}
\dot{\theta}_k(t)=\frac{1}{\hbar}\langle\phi_k|i\hbar\partial_t-\hat{H}|\phi_k\rangle~.
\end{equation}
The real function $\theta_k(t)$ is also known as the Lewis phase factor. Later we shall make use of eq.(\ref{3.6a}) to obtain the form of the Lewis phase and eventually the Berry phase for our model system. With the initial introduction regarding the invariant operator, we will try to find a suitable form of $\hat{I}(t)$ corresponding to the Hamiltonian in eq.(\ref{2.9}). 
\section{Lewis invariant of the system}\label{S4}
\noindent The form of the Hamiltonian in eq.(\ref{3.10}) suggests that the most general form of the Lewis invariant can be taken as
\begin{equation}\label{3.6}
\begin{split}
\hat{I}(t)=&\alpha_1(t)\hat{p}_1^2+\alpha_2(t)\hat{p}_2^2+\beta_1(t)\hat{x}_1^2+\beta_2(t)\hat{x}_2^2+\delta_1(t)(\hat{x}_1\hat{p}_1+\hat{p}_1\hat{x}_1)+\delta_2(t)(\hat{x}_2\hat{p}_2+\hat{p}_2\hat{x}_2)
\end{split}
\end{equation}
where $\alpha_1,\alpha_2,\beta_1,\beta_2,\delta_1$ and $\delta_2$ are all unknown time dependent parameters. The Lewis invariant method provides differential equations for these unknown parameters which yields the form of the invariant. In the following discussion, we shall develop the differential equations for these unknown time-dependent parameters. To proceed, we note that it is possible to break the invariant $\hat{I}(t)$ in eq.(\ref{3.6}) into two parts given by
\begin{equation}\label{3.12}
\hat{I}(t)=\hat{I}_1(t)+\hat{I}_2(t)
\end{equation}
where the forms of $\hat{I}_1$ and $\hat{I}_2(t)$ are given as follows
\begin{equation}\label{3.13}
\hat{I}_k(t)=\alpha_k(t)\hat{p}_k^2+\beta_k(t)\hat{x}_k^2+\delta_k(t)(\hat{x}_k\hat{p}_k+\hat{p}_k\hat{x}_k)~;~ k=1,2.
\end{equation}

\noindent
We shall now try to determine all the undetermined constants by inserting the form of $\hat{I}(t)$ from eq.(\ref{3.6}) back in eq.(\ref{3.1}) and equating the coefficients of all quadratic operators to zero. Computing $i\hbar\frac{\partial \hat{I}(t)}{\partial t}$ and $[\hat{I},\hat{H}]$, we obtain a set of ordinary differential equations given by
\begin{align}
&\dot{\alpha}_1-4\alpha_1d+2\delta_1a=0~,\label{3.7.1}\\
&\dot{\alpha}_2+4\alpha_2d+2\delta_2a=0~,\label{3.7.2}\\
&\dot{\beta}_1+4\beta_1d-2\delta_1b=0~,\label{3.7.3}\\
&\dot{\beta}_2-4\beta_2d-2\delta_2b=0~,\label{3.7.4}\\
&\dot{\delta}_1-\alpha_1b+\beta_1 a=0~,\label{3.7.5}\\
&\dot{\delta}_2-\alpha_2b+\beta_2 a=0~,\label{3.7.7}
\end{align}
We shall now introduce two time dependent parameters $\rho_1(t)$ and $\rho_2(t)$ such that
\begin{equation}\label{3.8}
\alpha_1(t)=\kappa_1\rho_1^2(t)~,~\alpha_2(t)=\kappa_2\rho_2^2(t)
\end{equation}
with $\kappa_1$ and $\kappa_2$ being two time independent unknown constants. Using the forms of $\alpha_1$ and $\alpha_2$, we can now obtain the form of the other unidentified time dependent parameters up to $\mathcal{O}(h)$ to be
\begin{align}
\beta_1(t)&=\frac{\kappa_1}{a^2}\left(\dot{\rho}_1^2+\rho_1\ddot{\rho}_1-2\rho_1^2\dot{d}-4\rho_1\dot{\rho}_1d+ab\rho_1^2\right)~,\label{3.9.1}\\
\beta_2(t)&=\frac{\kappa_2}{a^2}\left(\dot{\rho}_2^2+\rho_2\ddot{\rho}_2+2\rho_2^2\dot{d}+4\rho_2\dot{\rho}_2d+ab\rho_2^2\right)~,\label{3.9.2}\\
\delta_1(t)&=-\frac{\kappa_1}{a}(\rho_1\dot{\rho_1}-2\rho_1^2d)~,\label{3.9.3}\\
\delta_2(t)&=-\frac{\kappa_2}{a}(\rho_2\dot{\rho_2}+2\rho_2^2d)~,\label{3.9.4}
\end{align}
  To further simplify the forms of $\beta_1(t)$ and $\beta_2(t)$ from eq.(s)(\ref{3.9.1},\ref{3.9.2}), we substitute the forms of these two parameters (as well as $\delta_1(t)$ and $\delta_2(t)$) back in the following differential equations
\begin{align}
&\dot{\beta}_1+4\beta_1d-2\delta_1b=0~,\label{3.14.1}\\
&\dot{\beta}_2-4\beta_2d-2\delta_2b=0~.\label{3.14.2}
\end{align}  
We can finally obtain from the above two differential equations (using the forms of the parameters), two non-linear equations involving the parameters $\rho_1(t)$ and $\rho_2(t)$ as follows
\begin{align}
&\ddot{\rho}_1(t)+\left(ab(t)-2\dot{d}(t)\right)\rho_1(t)=\frac{a^2\xi_1^2}{\rho_1^3(t)}~,\label{3.15.1}\\
&\ddot{\rho}_2(t)+\left(ab(t)+2\dot{d}(t)\right)\rho_2(t)=\frac{a^2\xi_2^2}{\rho_2^3(t)}\label{3.15.2}
\end{align}
where $\xi_1^2$ and $\xi_2^2$ are integration constants. It is important to note that $\varepsilon_+^2+\varepsilon_\times^2=1$ and for a gravitational wave in the absence of the cross polarization term, $\epsilon_+=1$ which indicates $d(t)=\frac{1}{2}\dot{\chi}(t)$. Eq.(s)(\ref{3.15.1},\ref{3.15.2}) are the Ermakov-Pinney \cite{Ermakov_Pinney, Ermakov_Pinney_Final} equations corresponding to the two directions presented in this analysis. 

\noindent We shall now make use of the Ermakov-Pinney equations to subsequently simplify the forms of $\beta_1(t)$ and $\beta_2(t)$ from eq.(s)(\ref{3.9.1},\ref{3.9.2}) given as follows
\begin{align}
\beta_1(t)&=\frac{\kappa_1}{a^2}\left(\dot{\rho}_1^2-4\rho_1\dot{\rho}_1d\right)+\frac{\kappa_1\xi_1^2}{\rho_1^2}~,\label{3.16.1}\\
\beta_2(t)&=\frac{\kappa_2}{a^2}\left(\dot{\rho}_2^2+4\rho_2\dot{\rho}_2d\right)+\frac{\kappa_2\xi_2^2}{\rho_2^2}~.\label{3.16.2}
\end{align}
With the analytical forms of the parameters, we can now recast the separable Lewis invariants from eq.(\ref{3.13}) as follows
\begin{align}
\hat{I}_1(t)=&\kappa_1\rho_1^2\hat{p}_1^2+\left(m^2\kappa_1(\dot{\rho}_1^2-2\rho_1\dot{\rho_1}\dot{\chi}(t))+\frac{\kappa_1\xi_1^2}{\rho_1^2}\right)\hat{x}_1^2-m\kappa_1\rho_1(\dot{\rho}_1-\dot{\chi}(t)\rho_1)(\hat{x}_1\hat{p}_1+\hat{p}_1\hat{x}_1)~,\label{3.17.1}\\
\hat{I}_2(t)=&\kappa_2\rho_2^2\hat{p}_2^2+\left(m^2\kappa_2(\dot{\rho}_2^2+2\rho_2\dot{\rho_2}\dot{\chi}(t))+\frac{\kappa_2\xi_2^2}{\rho_2^2}\right)\hat{x}_2^2-m\kappa_2\rho_2(\dot{\rho}_2+\dot{\chi}(t)\rho_2)(\hat{x}_2\hat{p}_2+\hat{p}_2\hat{x}_2)~.\label{3.17.2}
\end{align}
It is important to note that $\kappa_1$ and $\kappa_2$ being arbitrary constants, can be set to unity for a simplified analysis. In order to proceed further, we would try to compute the raising and lowering operators corresponding to the new separable invariant operators $\hat{I}_1$ and $\hat{I}_2$.
\section{Obtaining the creation and annihilation operators}\label{S5}
\noindent In this section, we will obtain the creation and annihilation operators corresponding to the invariant operator $\hat{I}_1(t)$ using a simple ``\textit{completing the square}" approach and proceed to do the same for $\hat{I}_2(t)$. Our primary aim is to recast $\hat{I}_1(t)$ in such a way that $\hat{I}_1(t)=A_1^2(\hat{x}_1,\hat{p}_1)+A_2^2(\hat{x}_1,\hat{p}_1)+A_3$, where $A_3$ is a constant. We can recast $\hat{I}_1(t)$ in the following way
\begin{equation}\label{4.1}
\begin{split}
\frac{\hat{I}_1}{\kappa_1}=&\frac{\xi_1^2\hat{x}_1^2}{\rho_1^2}+\rho_1^2\hat{p}_1^2+m\rho_1(\rho_1\dot{\chi}(t)-\dot{\rho}_1)\hat{p}_1\hat{x}_1+m\rho_1(\rho_1\dot{\chi}(t)-\dot{\rho}_1)\hat{x}_1\hat{p}_1+m^2(\rho_1\dot{\chi}(t)-\dot{\rho}_1)^2\hat{x}_1^2-m^2(\rho_1\dot{\chi}(t)-\dot{\rho}_1)^2\hat{x}_1^2\\&+m^2(\dot{\rho}_1^2-2\rho_1\dot{\rho}_1\dot{\chi}(t))\hat{x}_1^2~.
\end{split}
\end{equation}
From the last line of the above equation, we can see that the final two terms cancel each other up to $\mathcal{O}(h)$. Hence, we can recast eq.(\ref{4.1}) in the following form
\begin{equation}\label{4.2}
\begin{split}
\frac{\hat{I}_1}{\kappa_1}=&\frac{\xi_1^2\hat{x}_1^2}{\rho_1^2}+\left(\rho_1\hat{p}_1+m(\rho_1\dot{\chi}(t)-\dot{\rho}_1)\hat{x}_1\right)^2~.
\end{split}
\end{equation}
From now on, we shall be working in the $\hbar=c=1$ unit. As a result, the commutation relation between the position and momentum variables takes the form $[\hat{x}_1,\hat{p}_1]=[\hat{x}_2,\hat{p}_2]=i$. We can now recast eq.(\ref{4.2}) in the following form
\begin{equation}\label{4.3}
\begin{split}
\hat{I}_1=&\kappa_1\left[\frac{\xi_1\hat{x}_1}{\rho_1}-i\left(\rho_1\hat{p}_1+m(\rho_1\dot{\chi}(t)-\dot{\rho}_1)\hat{x}_1\right)\right]\left[\frac{\xi_1\hat{x}_1}{\rho_1}+i\left(\rho_1\hat{p}_1+m(\rho_1\dot{\chi}(t)-\dot{\rho}_1)\hat{x}_1\right)\right]+\kappa_1\xi_1~.
\end{split}
\end{equation}
As $\kappa_1$ and $\xi_1$ are both arbitrary constants,  we can choose $\kappa_1=\frac{1}{2}$ and $\xi_1=1$. We can again recast eq.(\ref{4.3}) as follows
\begin{equation}\label{4.4}
\begin{split}
\hat{I}_1(t)=\hat{a}^\dagger_1(t)\hat{a}_1(t)+\frac{1}{2}
\end{split}
\end{equation}
where
\begin{equation}\label{4.5}
\begin{split}
\hat{a}_1&=\frac{1}{\sqrt{2}\rho_1}\left(\hat{x}_1+i\rho_1^2\hat{p}_1+im\rho_1(\rho_1\dot{\chi}-\dot{\rho}_1)\hat{x}_1\right)\\
&=\frac{1}{\sqrt{2}\rho_1}\left(\hat{x}_1+i\rho_1^2\hat{p}_1+im\rho_1(\rho_1\Gamma^1_{~01}-\dot{\rho}_1)\hat{x}_1\right)
\end{split}
\end{equation}
where $\Gamma^1_{~01}=\dot{\chi}(t)$. It can be easily checked using the form of eq.(\ref{4.5}) that $[\hat{a}_1,\hat{a}_1^\dagger]=1$. We can also recast the invariant operator $\hat{I}_2$ as follows
\begin{equation}\label{4.6}
\hat{I}_2(t)=\hat{a}_2^\dagger(t)\hat{a}_2(t)+\frac{1}{2}
\end{equation}
where
\begin{equation}\label{4.7}
\begin{split}
\hat{a}_2&=\frac{1}{\sqrt{2}\rho_2}\left(\hat{x}_2+i\rho_2^2\hat{p}_2-im\rho_2(\rho_2\dot{\chi}+\dot{\rho}_2)\hat{x}_2\right)\\
&=\frac{1}{\sqrt{2}\rho_2}\left(\hat{x}_2+i\rho_2^2\hat{p}_2+im\rho_2(\rho_2\Gamma^2_{~02}-\dot{\rho}_2)\hat{x}_2\right)
\end{split}
\end{equation}
where $\Gamma^2_{02}=-\dot{\chi}(t)$ and $[\hat{a}_2,\hat{a}_2^\dagger]=1$. From eq.(s)(\ref{4.5},\ref{4.7}), we can write down a generalized commutation relation involving the ladder operators given by
\begin{equation}\label{4.8}
[\hat{a}_j,\hat{a}_k^\dagger]=\delta_{jk}
\end{equation}
for $j,k=\{1,2\}$. With the forms of the creation and annihilation operators in hand, we shall now proceed to obtain the Lewis phase for the system.
\section{Lewis and Berry phase for the system}\label{S6}
\noindent In this section, we shall obtain the Lewis phase and eventually the Berry phase for the system. To obtain the Lewis phase, we need to properly construct the eigenstates corresponding to the Lewis invariant operators. It is important to note that the total Lewis invariant of the system can be divided into two Lewis invariants (eq.(\ref{3.12})) and each of the invariant operators can be expressed as a number operator plus a constant. Hence, it is straightforward to write down the total eigenstate corresponding to $\hat{I}(t)$ as a tensor product of two number states corresponding to each of the two directions given by
\begin{equation}\label{5.1}
|n_1,n_2\rangle=|n_1\rangle\otimes|n_2\rangle
\end{equation}
such that
\begin{equation}\label{5.2}
(\hat{a}_1\otimes\mathbb{1}_2)|0,n_2\rangle=0~,\text{and, }(\mathbb{1}_1\otimes\hat{a}_2)|n_1,0\rangle=0~.
\end{equation}
We can further define the action of the invariant operator on the number state as follows
\begin{equation}\label{5.3}
\begin{split}
\hat{I}(t)|n_1,n_2\rangle&=\left(\hat{I}_1(t)\otimes\mathbb{1}_2+\mathbb{1}_1\otimes\hat{I}_2(t)\right)|n_1,n_2\rangle\\
&=\left(n_1+\frac{1}{2}\right)|n_1,n_2\rangle+\left(n_2+\frac{1}{2}\right)|n_1,n_2\rangle\\
&=(n_1+n_2+1)|n_1,n_2\rangle~.
\end{split}
\end{equation}
It is straightforward to write down the number state as an action of creation operators on the individual vacuum states as follows
\begin{equation}\label{5.4}
|n_1,n_2\rangle=\frac{1}{\sqrt{n_1!n_2!}}\left((\hat{a}_1^\dagger)^{n_1}|0\rangle\right)\otimes\left((\hat{a}_2^\dagger)^{n_2}|0\rangle\right)~.
\end{equation}
For $\hbar=1$, we can recast eq.(\ref{3.6a}) (and for $|\phi_n\rangle=|n_1,n_2\rangle$) as follows
\begin{equation}\label{5.5}
\begin{split}
\dot{\theta}(t)=&\langle n_1,n_2|i(\partial_t)_1\otimes\mathbb{1}_2+\mathbb{1}_1\otimes i(\partial_t)_2-(\hat{H}_1(t)\otimes\mathbb{1}_2+\mathbb{1}_1\otimes\hat{H}_2(t))|n_1,n_2\rangle\\
=&\langle n_1|i(\partial_t)_1-\hat{H}_1(t)|n_1\rangle\langle n_2|\mathbb{1}_2|n_2\rangle+\langle n_1|\mathbb{1}_1|n_1\rangle\langle n_2|i(\partial_t)_2-\hat{H}_2(t)|n_2\rangle\\
=&\langle n_1|i(\partial_t)_1-\hat{H}_1(t)|n_1\rangle+\langle n_2|i(\partial_t)_2-\hat{H}_2(t)|n_2\rangle\\
=&\dot{\theta}_1(t)+\dot{\theta}_2(t)
\end{split}
\end{equation}
where
\begin{align}
\dot{\theta}_1(t)&=\langle n_1|i(\partial_t)_1-\hat{H}_1(t)|n_1\rangle\label{5.6a}~,\\
\dot{\theta}_2(t)&=\langle n_2|i(\partial_t)_2-\hat{H}_2(t)|n_2\rangle\label{5.6b}~.
\end{align}
We shall now make use of eq.(s)(\ref{5.6a},\ref{5.6b}) to obtain the forms of the Lewis phases $\theta_1(t)$ and $\theta_2(t)$. In order to obtain the form of $\dot{\theta}_1(t)$, we at first need to calculate the commutator bracket $[\hat{a}_1,i\partial_t-\hat{H}_1(t)]$ given as follows
\begin{equation}\label{5.7}
\begin{split}
[\hat{a}_1,i\partial_t-\hat{H}_1]|\psi\rangle=\left(-i\dot{\hat{a}}_1-[\hat{a}_1,\hat{H}_1]\right)|\psi\rangle~.
\end{split}
\end{equation}
The analytical form of $-i\dot{a}_1$ is given by
\begin{equation}\label{5.8}
-i\dot{\hat{a}}_1=\frac{i\dot{\rho}_1}{\sqrt{2}\rho_1^2}\hat{x}_1+\frac{\dot{\rho}_1\hat{p}_1}{\sqrt{2}}+\frac{m}{\sqrt{2}}\left[\dot{\rho}_1\Gamma^1_{~01}+\rho_1\dot{\Gamma}^1_{~01}-\ddot{\rho}_1\right]\hat{x}_1
\end{equation}
and the form of $[\hat{a}_1,\hat{H}_1]$ is given by
\begin{equation}\label{5.9}
\begin{split}
[\hat{a}_1,\hat{H}_1]&=\frac{i\hat{p}_1}{\sqrt{2}m\rho_1}+\frac{i\Gamma^1_{~01}\hat{x}_1}{\sqrt{2}\rho_1}+\frac{\rho_1m\omega_1^2\hat{x}_1}{\sqrt{2}}+\frac{\dot{\rho}_1\hat{p}_1}{\sqrt{2}}+\frac{m\Gamma^1_{~01}\dot{\rho}_1\hat{x}_1}{\sqrt{2}}~.
\end{split}
\end{equation}
Summing eq.(\ref{5.8}) with eq.(\ref{5.9}) and making use of the Ermakov-Pinney equation in eq.(\ref{3.15.1}) (for $\xi_1=1$ and $\dot{d}(t)=\frac{1}{2}\dot{\Gamma}^1_{~01}$), we can express the form of the commutator bracket in eq.(\ref{5.7}) as follows
\begin{equation}\label{5.10}
[\hat{a}_1,i\partial_t-\hat{H}_1]=-\frac{1}{m\rho_1^2}\hat{a}_1~.
\end{equation}
Making use of the analytical form of the commutator bracket given in eq.(\ref{5.10}), again we can recast the right hand side of eq.(\ref{5.6a}) as
\begin{equation}\label{5.11}
\begin{split}
\langle n_1|i(\partial_t)_1-\hat{H}_1(t)|n_1\rangle=-\frac{n_1}{m\rho_1^2}+\langle 0|i\partial_t-\hat{H}_1|0\rangle~.
\end{split}
\end{equation}
Note that the second term in the right hand side of the above equation is a constant arbitrary phase, and therefore can be chosen conveniently. A convenient form of the zero state contribution in our current analysis can be chosen as \cite{LewisRiesenfeld, Dittrich,Fring}\footnote{Note that the choice made here is inconsequential as we shall see in the subsequent discussion that this does not contribute to the Berry phase. }
\begin{equation}\label{5.12}
\langle 0|i\partial_t-\hat{H}_1|0\rangle=-\frac{1}{2m\rho_1^2}~.
\end{equation}
Using, eq.(s)(\ref{5.11},\ref{5.12}), we can write down the Lewis phase $\theta_1(t)$ as follows
\begin{equation}\label{5.13}
\theta_1(t)=-\left(n_1+\frac{1}{2}\right)\int_0^t\frac{d\tau}{m\rho_1^2(\tau)}~.
\end{equation}
Following a similar analysis, we can obtain the other Lewis phase $\theta_2(t)$ as follows
\begin{equation}\label{5.14}
\theta_2(t)=-\left(n_2+\frac{1}{2}\right)\int_0^t\frac{d\tau}{m\rho_2^2(\tau)}~.
\end{equation}
With the forms of $\theta_1(t)$ and $\theta_2(t)$ in hand, we can now proceed to obtain the geometric part of the phase making use of the adiabatic approximation. In the adiabatic approximation, $\ddot{\rho}_1(t)=\ddot{\rho}_2(t)=0$, and the modified Ermakov-Pinney equations corresponding to the two coordinates take the form
\begin{align}
\omega^2(t)-\dot{\Gamma}^1_{~01}(t)&\simeq\frac{1}{m^2\rho_1^4(t)}~,\label{5.15}\\
\omega^2(t)+\dot{\Gamma}^1_{~01}(t)&\simeq\frac{1}{m^2\rho_2^4(t)}~.\label{5.16}
\end{align}
Since $\dot{\Gamma}^1_{~01}(t)=\ddot{h}_{11}(t)/2$ is very small (as $h_{11}(t)$ is a very small quantity), it is evident that $\dot{\Gamma}^1_{~01}(t)<<\omega^2(t)$. We can therefore simplify eq.(s)(\ref{5.15},\ref{5.16}) to the following forms
\begin{align}
\frac{1}{m\rho_1^2(t)}&\simeq\omega(t)-\frac{\dot{\Gamma}^1_{~01}(t)}{2\omega(t)}~,\label{5.17}\\
\frac{1}{m\rho_2^2(t)}&\simeq\omega(t)+\frac{\dot{\Gamma}^1_{~01}(t)}{2\omega(t)}\label{5.18}~.
\end{align}
It is straightforward to show that, corresponding to the two coordinate directions, if one considers the harmonic oscillator frequencies to be different in the two coordinate directions then eq.(s)(\ref{5.17},\ref{5.18}) take the forms given by
\begin{align}
\frac{1}{m\rho_1^2(t)}&=\omega_1(t)-\frac{\dot{\Gamma}^1_{~01}(t)}{2\omega_1(t)}~,\frac{1}{m\rho_2^2(t)}=\omega_2(t)+\frac{\dot{\Gamma}^1_{~01}(t)}{2\omega_2(t)}\label{5.19}~.
\end{align}
Substituting eq.(\ref{5.19}) back in eq.(s)(\ref{5.13},\ref{5.14}), we obtain the following two relations
\begin{align}
\tilde{\theta}_1(t)&=-\left(n_1+\frac{1}{2}\right)\left(\int_0^t\omega_1(\tau)d\tau-\int_0^t\frac{\dot{\Gamma}^1_{~01}(\tau)}{2\omega_1(\tau)}d\tau\right)~,\label{5.20}\\
\tilde{\theta}_2(t)&=-\left(n_2+\frac{1}{2}\right)\left(\int_0^t\omega_2(\tau)d\tau+\int_0^t\frac{\dot{\Gamma}^1_{~01}(\tau)}{2\omega_2(\tau)}d\tau\right)\label{5.21}
\end{align}
where $\tilde{\theta}_n(t)$ (for $n=\{1,2\}$) denotes the Lewis phase $\theta_n(t)$ in the adiabatic approximation. From eq.(s)(\ref{5.20},\ref{5.21}), it is straightforward to observe that the first integrals introduce a dynamic phase whereas the second integrals are geometric in nature. Now we consider that the Hamiltonian of the system completes an adiabatic cycle at $t=\mathcal{T}$ in the parameter space and as a result it is possible to write down
\begin{equation}\label{5.21b}
\mathcal{R}(0)=\mathcal{R}(\mathcal{T})~;~~ \mathcal{R}=(b,d)~.
\end{equation} 
We can now easily write down the first order time derivative in terms of $\mathcal{R}$ as 
\begin{equation}\label{5.21c}
\frac{d}{dt}=\frac{d\mathcal{R}}{dt}\nabla_{\mathcal{R}}~.
\end{equation}
Hence, corresponding to the two coordinate directions and making use of eq.(\ref{5.21c}),  we can write down Berry's geometric phases to be of the following form 
\begin{align}
\Theta_1^{G}&=\left(n_1+\frac{1}{2}\right)\oint^\mathcal{R}\frac{1}{\omega_1} \vec{\nabla}_{\mathcal{R}}\left(\Gamma^1_{~01}\right).d\vec{\mathcal{R}}\label{5.22}~,\\
\Theta_2^{G}&=-\left(n_2+\frac{1}{2}\right)\oint^\mathcal{R}\frac{1}{\omega_2} \vec{\nabla}_{\mathcal{R}}\left(\Gamma^1_{~01}\right).d\vec{\mathcal{R}}\label{5.23}~.
\end{align}
Similar results for the Berry phases were obtained using a completely different method in \cite{Bibhas}. The presence of the Christoffel symbol $\Gamma^1_{~01}$ in the expressions for the Berry phases corresponding to the two spatial directions in the above two equations indicates the possibility that observing a non-zero Berry phase in this set up is equivalent to detecting a gravitational wave. We would like to make a remark now.  There is a possibility that a contribution to the Berry phase may also arise from the boundary term (eq.(\ref{2.9a})). In section \ref{S7b}, we shall calculate this contribution explicitly and show that it vanishes. 
\section{Explicit Berry phase calculation}\label{S7}
\noindent In this section, we shall make use of eq.(s)(\ref{5.22},\ref{5.23}) to compute some explicit forms of the Berry phases corresponding to the two coordinate directions. 

\noindent For a linearly polarized gravitational wave 
\begin{equation}\label{6.01}
h_{jk}(t)=2f_0\cos\Omega t\left(\varepsilon_+\sigma^3_{jk}+\varepsilon_{\times}\sigma^1_{jk}\right)
\end{equation}
where we have set $\chi(t)=f_0\cos\Omega t$ in eq.(\ref{2.3}) with $f_0$ being the amplitude of the gravitational wave and $\Omega$ being the frequency of the same. It is important to note that we have considered that the gravitational wave is carrying plus polarization only. Using eq.(\ref{6.01}), we can write down the first order time derivative of the Christoffel symbol $\dot{\Gamma}^1_{~01}=\frac{\ddot{h}_{11}(t)}{2}=-f_0\Omega^2\varepsilon_+\cos\Omega t$. For a linearly polarized gravitational wave, eq.(s)(\ref{5.22},\ref{5.23}) give
\begin{align}
\Theta_1^G&=-f_0\Omega^2\varepsilon_+\left(n_1+\frac{1}{2}\right)\int_0^\frac{2\pi}{\Omega} d\tau\frac{\cos\Omega\tau}{2\omega_1(\tau)}\label{6.02}~,\\
\Theta^G_2&=f_0\Omega^2\varepsilon_+\left(n_2+\frac{1}{2}\right)\int_0^\frac{2\pi}{\Omega} d\tau\frac{\cos\Omega\tau}{2\omega_2(\tau)}\label{6.03}~.
\end{align}
Now for an explicit set of choices of the frequencies $\omega_1(t)=\omega_{01}\cos\Omega t$ and $\omega_2(t)=\omega_{02}\cos\Omega(t)$, we can evaluate $\Theta_1^G$ and $\Theta_2^G$ as follows
\begin{align}
\Theta_1^G&=-\frac{f_0\pi\Omega\varepsilon_+}{\omega_{01}}\left(n_1+\frac{1}{2}\right)~,\label{6.04}\\\Theta_2^G&=\frac{f_0\pi\Omega\varepsilon_+}{\omega_{02}}\left(n_2+\frac{1}{2}\right)~.\label{6.05}
\end{align}
Our second choice of the frequencies are $\omega_1(t)=\omega_{1}+\tilde{\omega}_{1}\cos\Omega t$ and $\omega_2(t)=\omega_{2}+\tilde{\omega}_{2}\cos\Omega t$, such that $|\omega_{1}|\neq|\tilde{\omega}_{1}|$ and  $|\omega_{2}|\neq|\tilde{\omega}_{2}|$. Under this choice of frequencies, we can evaluate $\Theta_1^G$ and $\Theta_2^G$ given by
\begin{align}
\Theta_1^G=&-\frac{\pi f_0\Omega\varepsilon_+}{\tilde{\omega}_1}\left(n_1+\frac{1}{2}\right)~,\label{6.06}\\
\Theta_2^G=&\frac{\pi f_0\Omega\varepsilon_+}{\tilde{\omega}_2}\left(n_2+\frac{1}{2}\right)~.\label{6.07}
\end{align}
\section{Geometric phase from the boundary term}\label{S7b}
In this section, we shall calculate the geometric phase from the boundary term in eq.(\ref{2.9a}). We start by rewriting the term (eq.(\ref{2.9a})) as
\begin{equation}\label{7.1}
\begin{split}
\mathcal{B}&=\frac{m}{2}\int_0^{\mathcal{T}}d\tau \frac{d}{d\tau}\left(\Gamma^{j}_{~0k}(\tau)x_j(\tau)x^k(\tau)\right)\\
&=\frac{m}{2}\oint^{\mathcal{R}'}d\vec{\mathcal{R}}'.\vec{\nabla}_{\mathcal{R}'}\left(\Gamma^{j}_{~0k}(\tau)x_j(\tau)x^k(\tau)\right)
\end{split}
\end{equation}
where $\mathcal{R}'(0)=\mathcal{R}'(\mathcal{T})$. Eq.(\ref{7.1}) reveals that $\mathcal{B}$ is geometric in nature. We can now raise the phasespace position variables to operator status and after a little bit of calculation, we can recast eq.(\ref{7.1}) in the following form (where the gravitational wave is considered to be carrying plus polarization only)
\begin{equation}\label{7.2}
\begin{split}
\hat{\mathcal{B}}&=\frac{m}{2}\left\{\Gamma^1_{~01}(\tau)\hat{x}_1^2(\tau)\otimes\hat{\mathbb{1}}_{2}+\Gamma^2_{~02}(\tau)~\hat{\mathbb{1}}_{1}\otimes\hat{x}_2^2(\tau)\right\}\Bigr\rvert_{\tau=0}^{\tau=\mathcal{T}}\\
&=\frac{m}{4}\dot{h}_+(\tau)\left(\hat{x}_1^2(\tau)\otimes\hat{\mathbb{1}}_{2}-\hat{\mathbb{1}}_{1}\otimes\hat{x}_2^2(\tau)\right)\Bigr\rvert_{\tau=0}^{\tau=\mathcal{T}}\\
&=\hat{\mathcal{B}}_1\otimes\hat{\mathbb{1}}_{2}+\hat{\mathbb{1}}_{1}\otimes\hat{\mathcal{B}}_2
\end{split}
\end{equation}
with the definitions $\hat{\mathcal{B}}_1\equiv\frac{m}{4}\left(\dot{h}_+(\mathcal{T})\hat{x}_1^2(\mathcal{T})-\dot{h}_+(0)\hat{x}_1^2(0)\right)$ and $\hat{\mathcal{B}}_2\equiv-\frac{m}{4}\left(\dot{h}_+(\mathcal{T})\hat{x}_2^2(\mathcal{T})-\dot{h}_+(0)\hat{x}_2^2(0)\right)$. The total geometric phase can be obtained by calculating $\langle n_1|\hat{\mathcal{B}}_1|n_1\rangle$ and $\langle n_2|\hat{\mathcal{B}}_2|n_2\rangle$ separately and adding them. Before proceeding further, we need to obtain $\hat{x}_1^2(\tau)$ and $\hat{x}_2^2(\tau)$ in terms of the corresponding raising and lowering operators.  Using the form of the lowering operator in the first coordinate direction from eq.(\ref{4.5}) and making use of its complex conjugate,  we obtain
\begin{equation}\label{7.3}
\begin{split}
\hat{x}_1(\tau)=\frac{\rho_1(\tau)}{\sqrt{2}}\left(\hat{a}_1+\hat{a}_1^\dagger\right)\implies \hat{x}_1^2(\tau)=\frac{\rho_1^2(\tau)}{2}\left(\hat{a}_1^2+\hat{a}_1^{\dagger2}+2\hat{a}_1^\dagger\hat{a}_1+1\right)~.
\end{split}
\end{equation}
Similarly by using eq.(\ref{4.7}) and its complex conjugate, we obtain
\begin{equation}\label{7.4}
\begin{split}
\hat{x}_2(\tau)=\frac{\rho_2(\tau)}{\sqrt{2}}\left(\hat{a}_2+\hat{a}_2^\dagger\right)\implies \hat{x}_2^2(\tau)=\frac{\rho_2^2(\tau)}{2}\left(\hat{a}_2^2+\hat{a}_2^{\dagger2}+2\hat{a}_2^\dagger\hat{a}_2+1\right)~.
\end{split}
\end{equation}
We shall now compute the contribution of the boundary term towards the geometric phase in the first coordinate direction as follows
\begin{equation}\label{7.5}
\begin{split}
\langle n_1|\hat{\mathcal{B}}_1|n_1\rangle&=\frac{1}{2}\langle n_1|\frac{m}{4}\left(\dot{h}_+(\mathcal{T})\rho_1^2(\mathcal{T})-\dot{h}_+(0)\rho_1^2(0)\right)\left(\hat{a}_1^2+\hat{a}_1^{\dagger2}+2\hat{a}_1^\dagger\hat{a}_1+1\right)|n_1\rangle\\
&=\frac{m}{4}\left(n_1+\frac{1}{2}\right)\left(\dot{h}_+(\mathcal{T})\rho_1^2(\mathcal{T})-\dot{h}_+(0)\rho_1^2(0)\right)~.
\end{split}
\end{equation}
Our next aim is to substitute the form of $\rho_1^2(\tau)$ in the above equation. Using the Ermakov-Pinney equation (eq.(\ref{5.17})) under the adiabatic approximation (with $\omega(\tau)$ replaced by $\omega_1(\tau)$), we have
\begin{equation}\label{7.6}
\begin{split}
m&\rho_1^2(\tau)=\frac{1}{\omega_1(\tau)-\frac{\dot{\Gamma}^1_{~01}(\tau)}{2\omega_1(\tau)}}\\
\implies &\rho_1^2(\tau)\simeq \frac{1}{m \omega_1(\tau)}+\frac{\ddot{h}_+(\tau)}{4m\omega_1^3(\tau)}~.
\end{split}
\end{equation}
Substituting eq.(\ref{7.6}) back in eq.(\ref{7.5}) and keeping terms upto $\mathcal{O}(h)$, we obtain
\begin{equation}\label{7.7}
\langle n_1|\hat{\mathcal{B}}_1|n_1\rangle\simeq\frac{1}{4}\left(n_1+\frac{1}{2}\right)\left(\frac{\dot{h}_+(\mathcal{T})}{\omega_1(\mathcal{T})}-\frac{\dot{h}_+(0)}{\omega_1(0)}\right)~.
\end{equation}
Similarly, the term $\langle n_2|\hat{B}_2|n_2\rangle$ can be calculated and is given by
\begin{equation}\label{7.8}
\langle n_2|\hat{\mathcal{B}}_2|n_2\rangle\simeq-\frac{1}{4}\left(n_2+\frac{1}{2}\right)\left(\frac{\dot{h}_+(\mathcal{T})}{\omega_2(\mathcal{T})}-\frac{\dot{h}_+(0)}{\omega_2(0)}\right)~.
\end{equation}
Hence, the total geometric phase due the boundary term in an anisotropic harmonic oscillator reads
\begin{equation}\label{7.9}
\begin{split}
\langle n_1,n_2|\hat{\mathcal{B}}|n_1,n_2\rangle&=\langle n_1|\hat{\mathcal{B}}_1|n_1\rangle\langle n_2|\hat{\mathbb{1}}_2|n_2\rangle+\langle n_1|\hat{\mathbb{1}}_1|n_1\rangle\langle n_2|\hat{\mathcal{B}}_2|n_2\rangle\\
&=\langle n_1|\hat{\mathcal{B}}_1|n_1\rangle+\langle n_2|\hat{\mathcal{B}}_2|n_2\rangle\\
&=\frac{1}{4}\left(n_1+\frac{1}{2}\right)\left(\frac{\dot{h}_+(\mathcal{T})}{\omega_1(\mathcal{T})}-\frac{\dot{h}_+(0)}{\omega_1(0)}\right)-\frac{1}{4}\left(n_2+\frac{1}{2}\right)\left(\frac{\dot{h}_+(\mathcal{T})}{\omega_2(\mathcal{T})}-\frac{\dot{h}_+(0)}{\omega_2(0)}\right)~.
\end{split}
\end{equation}
For an isotropic harmonic oscillator with the oscillator frequency $\omega_1(\tau)=\omega_2(\tau)
=\omega(\tau)$, interacting with a gravitational wave, we obtain
\begin{equation}\label{7.10}
\langle n|\hat{B}|n\rangle=\frac{1}{4}\left(n_1-n_2\right)\left(\frac{\dot{h}_+(\mathcal{T})}{\omega(\mathcal{T})}-\frac{\dot{h}_+(0)}{\omega(0)}\right)
\end{equation}
where $|n\rangle=|n_1,n_2\rangle=|n_1\rangle\otimes |n_2\rangle$. Both $h(\tau)$ and $\omega(\tau)$ belongs to the parameter space $\mathcal{R}'$ and we therefore know, $h(\mathcal{T})=h(0)$ and $\omega(\mathcal{T})=\omega(0)$. For example, the linearly polarized gravitational wave from eq.(\ref{6.01}) has the component $h_+(\tau)=2f_0\cos\Omega \tau$. Here, $\mathcal{T}=\frac{2\pi}{\Omega}$ and $h_+(2\pi/\Omega)=h_+(0)$. Therefore, from eq.(\ref{7.9}) (anisotropic harmonic oscillator) and eq.(\ref{7.10}) (isotropic harmonic oscillator), we find that the contribution to the geometric phase coming from the boundary term (eq.(\ref{2.9a})) vanishes. 
\section{Incident gravitational wave with cross polarization only}\label{S8a}
In this section, we shall consider a gravitational wave incident on a two dimensional isotropic harmonic oscillator with cross polarization only ($\varepsilon_+=0$). The Hamiltonian operator of the system can be expressed as
\begin{equation}\label{8.1}
\begin{split}
\hat{H}(t)&=\frac{a(t)}{2}\left(\hat{p}_1^2+\hat{p}_2^2\right)+\frac{b(t)}{2}(\hat{x}_1^2+\hat{x}_2^2)+f(t)(\hat{x}_1\hat{p}_2+\hat{p}_1\hat{x}_2)\\
&=\frac{a(t)}{2}\left(\hat{p}_1^2+\hat{p}_2^2\right)+\frac{b(t)}{2}(\hat{x}_1^2+\hat{x}_2^2)+\frac{f(t)}{2}(\hat{x}_1\hat{p}_2+\hat{p}_2\hat{x}_1)+\frac{f(t)}{2}(\hat{p}_1\hat{x}_2+\hat{x}_2\hat{p}_1)
\end{split}
\end{equation}
where $a(t)=\frac{1}{m}$, $b(t)=m\omega^2(t)$, and $f(t)=\varepsilon_\times\dot{\chi}(t)$ with $\chi(t)$ denoting the amplitude of the gravitational wave as before. To write down the last line in eq.(\ref{8.1}), we have made use of the commutation relation $[\hat{x}_i,\hat{p}_j]=i\hbar\delta_{ij}$ ($i,j=1,2$).  Before trying to obtain a Lewis invariant corresponding to the harmonic oscillator-gravitational wave system, we make a change of coordinates as follows
\begin{equation}\label{8.2}
\hat{x}_+\equiv\frac{1}{\sqrt{2}}\left(\hat{x}_1+\hat{x}_2\right)~,~~\hat{x}_-\equiv\frac{1}{\sqrt{2}}\left(\hat{x}_1-\hat{x}_2\right)~.
\end{equation}
Corresponding to these new coordinate directions, one can now define two new momentum operators in the following way
\begin{equation}\label{8.3}
\begin{split}
\hat{p}_+\equiv\frac{1}{\sqrt{2}}\left(\hat{p}_1+\hat{p}_2\right)~,~~\hat{p}_-\equiv\frac{1}{\sqrt{2}}\left(\hat{p}_1-\hat{p}_2\right)~.
\end{split}
\end{equation}
It is easy to check the following set of commutation relations among the new position and momentum operators,
\begin{equation}\label{8.4}
\begin{split}
[\hat{x}_+,\hat{p}_+]=[\hat{x}_-,\hat{p}_-]=i\hbar~,~~[\hat{x}_+,\hat{x}_-]=[\hat{p}_+,\hat{p}_-]=[\hat{x}_+,\hat{p}_-]=[\hat{x}_-,\hat{p}_+]=0~.
\end{split}
\end{equation}
We can therefore claim that $\hat{p}_+$ is the momentum conjugate to $\hat{x}_+$ and $\hat{p}_-$ is the momentum conjugate to $\hat{x}_-$. In terms of these newly defined operators, we can write down the old operators as
\begin{equation}\label{8.5}
\hat{x}_1=\frac{1}{\sqrt{2}}(\hat{x}_++\hat{x}_-)~,~~\hat{x}_2=\frac{1}{\sqrt{2}}(\hat{x}_+-\hat{x}_-)~,~~\hat{p}_1=\frac{1}{\sqrt{2}}(\hat{p}_++\hat{p}_-)~,~\text{and}~~\hat{p}_2=\frac{1}{\sqrt{2}}(\hat{p}_+-\hat{p}_-)~.
\end{equation}
Making use of the above equation, one can rewrite the Hamiltonian in eq.(\ref{8.1}) in terms of the new ``$\pm$" operators as follows
\begin{equation}\label{8.6}
\begin{split}
\hat{H}(t)&=\frac{a}{2}\left(\hat{p}_+^2+\hat{p}_-^2\right)+\frac{b(t)}{2}\left(\hat{x}_+^2+\hat{x}_-^2\right)+\frac{f(t)}{2}(\hat{x}_+\hat{p}_++\hat{p}_+\hat{x}_+)-\frac{f(t)}{2}(\hat{x}_-\hat{p}_-+\hat{p}_-\hat{x}_-)\\
&=\frac{a}{2}\hat{p}_+^2+\frac{b(t)}{2}\hat{x}_+^2+\frac{f(t)}{2}(\hat{x}_+\hat{p}_++\hat{p}_+\hat{x}_+)+\frac{a}{2}\hat{p}_-^2+\frac{b(t)}{2}\hat{x}_-^2-\frac{f(t)}{2}(\hat{x}_-\hat{p}_-+\hat{p}_-\hat{x}_-)\\
&=\hat{H}_+(t)+\hat{H}_-(t)
\end{split}
\end{equation}
where $\hat{H}_\pm(t)=\frac{a}{2}\hat{p}_\pm^2+\frac{b(t)}{2}\hat{x}_\pm^2\pm\frac{f(t)}{2}(\hat{x}_\pm\hat{p}_\pm+\hat{p}_\pm\hat{x}_\pm)$. It is very interesting to see that by applying the coordinate changes introduced in eq.(\ref{8.2}) (and the conjugate momentum operators in eq.(\ref{8.3})), the coupled Hamiltonian in the $\{1,2\}$ basis now gets decoupled. The reason behind this decoupling can be easily understood via a simple diagram.
\begin{figure}
\begin{center}
\includegraphics[scale=0.25]{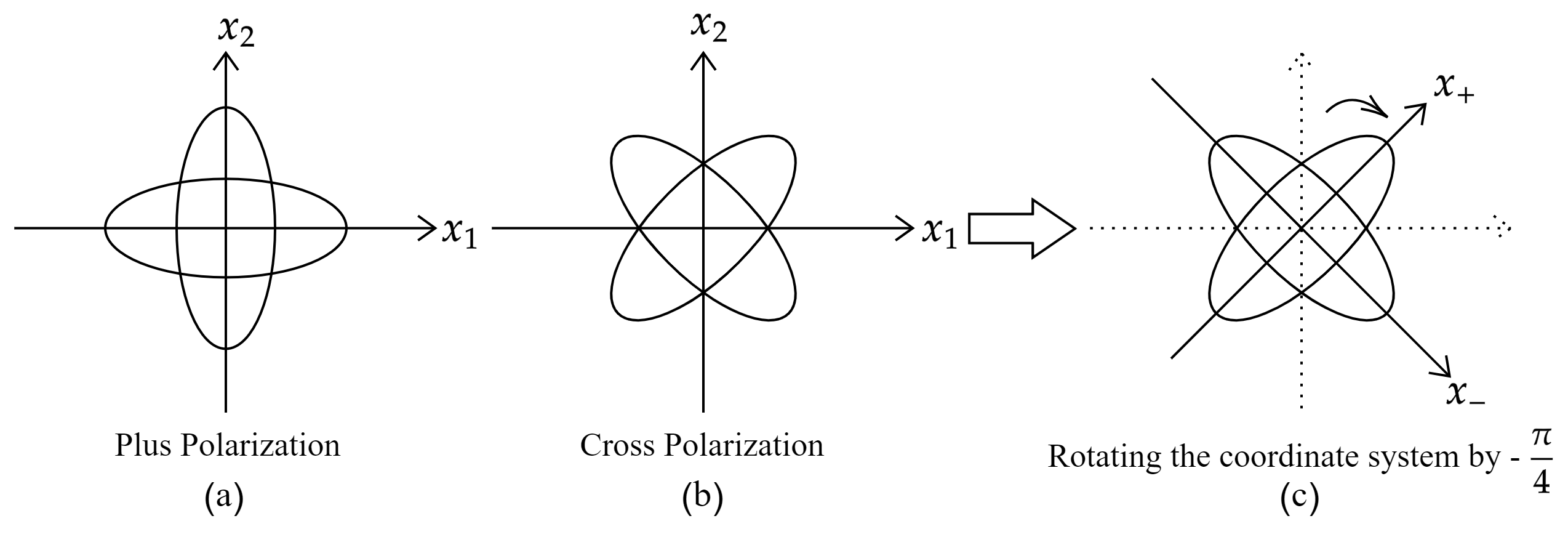}
\caption{In the new $`\pm'$ coordinate basis, the cross polarization effectively behaves like the plus polarization and therefore it is possible to decouple the Hamiltonian in this new coordinate system.}\label{F1}
\end{center}
\end{figure}
In Figure \ref{F1}, \ref{F1}(a) denotes a time slice of an incoming gravitational wave carrying a plus polarization only in the $\{x_1,x_2\}$ coordinate system. Figure \ref{F1}(b) denotes an incident gravitational wave with cross polarization only in the same coordinate basis. Next, we rotate the coordinate system by an angle $\pi/4$ in the clockwise direction and the new coordinate axes are now annotated as $x_+$ and $x_-$. As can be seen from Figure \ref{F1}(c), in this new coordinate system the cross polarization effectively behaves as the plus polarization. This is the primary reason behind the decoupling of the Hamiltonian as can be seen from eq.(\ref{8.6}) after implementing the coordinate changes implemented in eq.(\ref{8.2}). 

\noindent Our next aim is to obtain the Lewis invariant corresponding to the Hamiltonian given in eq.(\ref{8.6}). Due to the striking structural similarities among the Hamiltonian in eq.(\ref{8.6}) with that of the Hamiltonian for the plus polarization case in eq.(\ref{3.10}), we can proceed to obtain the Lewis phases as before. As before, one can propose the Lewis invariants corresponding to the $x_+$ and $x_-$ coordinate directions as
\begin{equation}\label{8.7}
\hat{I}_\pm(t)=\alpha_{\pm}(t)\hat{p}_\pm^2+\beta_\pm(t)\hat{x}_\pm^2+\delta_{\pm}(t)(\hat{x}_\pm\hat{p}_{\pm}+\hat{p}_\pm\hat{x}_\pm)~.
\end{equation}
We introduce two time dependent parameters $\rho_+(t)$ and $\rho_-(t)$ as follows
\begin{equation}\label{8.8}
\alpha_\pm(t)=\kappa_\pm\rho_\pm^2(t)~.
\end{equation}
Calculating $i\hbar\frac{\partial \hat{I}_\pm(t)}{\partial t}+[\hat{I}_\pm(t),\hat{H}_{\pm}(t)]=0$ for the proposed forms of the two invariants in eq.(\ref{8.7}), we obtain six ordinary differential equations. Making use of eq.(\ref{8.8}) in these set of differential equations, we eventually obtain two Ermakov-Pinney equations as
\begin{align}
\ddot{\rho}_+(t)+\left(ab(t)-\dot{f}(t)\right)\rho_+(t)&=\frac{a^2\xi_+^2}{\rho_+^3(t)}~,\label{8.9a}\\
\ddot{\rho}_-(t)+\left(ab(t)+\dot{f}(t)\right)\rho_-(t)&=\frac{a^2\xi_-^2}{\rho_-^3(t)}\label{8.9b}
\end{align}
where $\xi_+^2$ and $\xi_-^2$ are the two integration constants. Using the above two Ermakov-Pinney equations, we can obtain the form of the parameters as
\begin{align}
\beta_+=\frac{\kappa_+}{a^2}\left(\dot{\rho}_+^2-2\rho_+\dot{\rho}_+f(t)\right)+\frac{\kappa_+\xi_+^2}{\rho_+^2}~&,~\beta_-=\frac{\kappa_-}{a^2}\left(\dot{\rho}_-^2+2\rho_-\dot{\rho}_-f(t)\right)+\frac{\kappa_-\xi_-^2}{\rho_-^2}~,\label{9.0a}\\
\delta_+(t)=-\frac{\kappa_+}{a^2}\left(\rho_+\dot{\rho}_+-\rho_+^2f(t)\right)~&,~\delta_-(t)=-\frac{\kappa_-}{a^2}\left(\rho_-\dot{\rho}_-+\rho_-^2f(t)\right)~.\label{9.0b}
\end{align}
Using eq.(s)(\ref{8.8},\ref{9.0a},\ref{9.0b}), we can rewrite the Lewis invariants as follows
\begin{align}
\hat{I}_+(t)&=\kappa_+\rho_+\hat{p}_+^2+\left(\kappa_+m^2\left(\dot{\rho}_+^2-2\varepsilon_\times\dot{\chi}(t)\rho_+\dot{\rho}_+\right)+\frac{\kappa_+\xi_+^2}{\rho_+^2}\right)\hat{x}_+^2-m\kappa_+\rho_+\left(\dot{\rho}_+-\varepsilon_\times\dot{\chi}(t)\rho_+\right)\left(\hat{x}_+\hat{p}_++\hat{p}_+\hat{x}_+\right)\label{9.1a}~,\\
\hat{I}_-(t)&=\kappa_-\rho_-\hat{p}_-^2+\left(\kappa_-m^2\left(\dot{\rho}_-^2+2\varepsilon_\times\dot{\chi}(t)\rho_-\dot{\rho}_-\right)+\frac{\kappa_-\xi_-^2}{\rho_-^2}\right)\hat{x}_-^2-m\kappa_-\rho_-\left(\dot{\rho}_-+\varepsilon_\times\dot{\chi}(t)\rho_-\right)\left(\hat{x}_-\hat{p}_-+\hat{p}_-\hat{x}_-\right)\label{9.1b}~.
\end{align}
Making use of the completing the square approach, one can now obtain the following annihilation operators corresponding to the $``\pm"$ coordinate directions as
\begin{align}
\hat{a}_+&=\frac{1}{\sqrt{2}\rho_+}\left(\hat{x}_++i\rho_+^2\hat{p}_++im\rho_+\left(\varepsilon_\times\dot{\chi}\rho_+-\dot{\rho}_+\right)\hat{x}_+\right)~,\label{9.2a}\\
\hat{a}_-&=\frac{1}{\sqrt{2}\rho_-}\left(\hat{x}_-+i\rho_-^2\hat{p}_--im\rho_-\left(\varepsilon_\times\dot{\chi}\rho_-+\dot{\rho}_-\right)\hat{x}_-\right)~.
\end{align}
Again it is straightforward to check that $[\hat{a}_A,\hat{a}^\dagger_B]=\delta_{AB}$  and $[\hat{a}_A,\hat{a}_B]=[\hat{a}^\dagger_A,\hat{a}^\dagger_B]=0$ where $A,B=+,-$. Instead of going through the same procedure, it is possible to write down the Lewis phases, corresponding to the two coordinate directions as follows
\begin{align}
\theta_+(t)=-(n_++\frac{1}{2})\int_0^t\frac{d\tau}{m\rho_+^2(\tau)}\label{9.3a}~,\\
\theta_-(t)=-(n_-+\frac{1}{2})\int_0^t\frac{d\tau}{m\rho_-^2(\tau)}\label{9.3b}
\end{align}
where $n_+$ and $n_-$ denotes the eigenvalues of the number operators $\hat{N}_+=\hat{a}^\dagger_+\hat{a}_+$  and $\hat{N}_-=\hat{a}^\dagger_-\hat{a}_-$ corresponding to the eigenstate $|n_+,n_-\rangle=|n_+\rangle\otimes|n_-\rangle$. Here the number state of the system is expressed as a tensor product state of the individual number states.
Using the adiabatic approximation in eq.(s)(\ref{8.9a},\ref{8.9b}), we can recast the above Lewis phases as
\begin{align}
\theta_+(t)&\simeq-\left(n_++\frac{1}{2}\right)\int_0^t\omega(\tau) d\tau+\left(n_++\frac{1}{2}\right)\int_0^t d\tau\frac{\varepsilon_\times\ddot{\chi}(\tau)}{2\omega(\tau)^2}~,\label{9.4a}\\
\theta_-(t)&\simeq-\left(n_-+\frac{1}{2}\right)\int_0^t\omega(\tau) d\tau-\left(n_-+\frac{1}{2}\right)\int_0^t d\tau\frac{\varepsilon_\times\ddot{\chi}(\tau)}{2\omega(\tau)^2}~.\label{9.4b}
\end{align}
The second integrals in eq.(s)(\ref{9.4a},\ref{9.4b}) are geometric in nature. Therefore, corresponding to the two new coordinate directions, we can write down Berry's geometric phases as
\begin{align}
\Theta^G_+&=\left(n_++\frac{1}{2}\right)\oint^\mathcal{R}\frac{1}{\omega}\vec{\nabla}_\mathcal{R}\left(\Gamma^1_{~02}\right).d\vec{\mathcal{R}}~,\label{9.5a}\\
\Theta^G_-&=-\left(n_-+\frac{1}{2}\right)\oint^\mathcal{R}\frac{1}{\omega}\vec{\nabla}_\mathcal{R}\left(\Gamma^1_{~02}\right).d\vec{\mathcal{R}}\label{9.5b}
\end{align} 
where $\mathcal{R}(0)=\mathcal{R}(\mathcal{T})$ and $\mathcal{R}=(b,f)$. Using eq.(s)(\ref{9.5a},\ref{9.5b}), one can now write down the total Berry phase of the system as
\begin{equation}\label{9.6}
\Theta^G=\left(n_+-n_-\right)\oint^\mathcal{R}\frac{1}{\omega}\vec{\nabla}_\mathcal{R}\left(\Gamma^1_{~02}\right).d\vec{\mathcal{R}}~.
\end{equation} 
It is quite easily understandable that the total Berry phase of the system is independent of the choice of the coordinate basis. Contrary to the claim made in \cite{Bibhas}, we do observe that for different values of $n_+$ and $n_-$, there exists a Berry phase which is dependent on the cross polarization of the gravitational wave. Hence, not only the plus polarization, the cross polarization also carries signatures of a geometric phase.
\section{Summary}\label{S8}
\noindent In this work we have considered a gravitational wave interacting with an isotropic two dimensional harmonic oscillator. Starting from the traditional representation of the gravitational wave (in the transverse traceless gauge) to be in both the plus and cross polarizations, we proceed by omitting the cross polarization terms. The reason behind dropping the cross polarization is to avoid unnecessary complications during the calculation. We have then used the Lewis invariant method to obtain a suitable Lewis invariant corresponding to the total Hamiltonian of the system. Our main goal is to obtain a Berry phase from the Lewis phase by making an adiabatic approximation. We follow essentially the approach in  \cite{Dittrich} in achieving this task. In order to find a Lewis invariant, we have used an ansatz considering contributions from all quadratic order terms in the phase space variables and finally obtained two separable Lewis invariants. Using the ``\textit{completing the square}" approach, we have then obtained the ladder operators corresponding to the two coordinate directions from the obtained Lewis invariants. We then make use of the form of the ladder operator to obtain a commutation relation between the ladder operator and $i\partial_t-\hat{H}_k$ operator, $k=1,2$. Using this commutation relation, we finally obtain the analytical form of $\langle n_k|i(\partial_t)_k-\hat{H}_k|n_k\rangle$ which is essentially equal to the total time derivative of the Lewis phase $\theta_k$ ($k=1,2$). Our next aim was to extract the geometric part of the phase in the Lewis phase itself.  In order to do so we have made use of the Ermakov Pinney equations corresponding to the two spatial directions and made use of the adiabatic approximation to obtain the geometric part of the Lewis phase. Under this approximation, the Lewis phase breaks into two separate parts. The first part is dynamic in nature whereas the second part, which is completely dependent on the gravitational perturbation due to the incoming gravitational wave, is geometrical in nature. These are the Berry phases which arise solely due to the passing of the gravitational wave. In order to truly underpin the underlying nature of Berry's geometric phase we have made use of the linearly polarized gravitational wave template and made use of explicit values of the frequencies of the harmonic oscillator system. The Berry phase corresponding to the two spatial directions entirely depends upon the amplitude of the linearly polarized gravitational wave, amplitude of the plus polarization, and frequency of the incident gravitational wave. The approach that we have taken to obtain the Berry phase is based on finding the Lewis invariant and Lewis phase and then making an adiabatic approximation. This approach differs from the one in \cite{Bibhas} where similar results for the Berry phase were obtained using a ladder operator approach. The main objective of our study is to show that a non-trivial Berry phase in the setup that we have considered, can prove to be an effective tool for the detection of gravitational waves. In this regard, we have also calculated the contribution of the boundary term towards the total geometric phase of the system for both the isotropic and anisotropic harmonic oscillator. Now due to the periodicity of the parameters belonging to the parameter space, we observe that during a complete period of revolution the boundary term picks up no geometric phase for the well known gravitational wave templates. Finally, we consider a gravitational wave carrying cross polarization only. For this analysis, we have considered a gravitational wave carrying cross polarization only with an isotropic two dimensional harmonic oscillator. In order to calculate the Lewis phases for the new harmonic oscillator-gravitational wave system, we have made use of a rotated coordinate system such that the cross polarization is effectively similar to the plus polarization in this basis. Such a rotated coordinate system helps us to decouple the Hamiltonian and progress similarly to the plus polarization case. We then obtain the Lewis phases corresponding to the two new coordinate directions and eventually arrive at the geometric Berry phases by applying the adiabatic approximation. The total Berry phase of the system does not depend upon the choice of the coordinate basis and as can be observed from eq.(\ref{9.6}), it is dependent upon the cross polarization of the gravitational wave. This is quite contrary to the claim made in \cite{Bibhas} which states that the Berry phase is generated entirely by the plus polarization of the gravitational wave and does not depend upon the cross polarization of the gravitational wave. Therefore, our analysis with a gravitational wave carrying the cross polarization only is not in accord with the claim made in \cite{Bibhas}.

\section*{Appendix: Form of the coefficients in the Lewis invariant when the gravitational wave consists of both the plus and cross polarization terms}
In this appendix, we try to obtain a Lewis invariant considering the Hamiltonian in eq.\ref{2.9} which includes the interaction of the gravitational wave with the two dimensional harmonic oscilltaor carrying both the plus and cross polarization. Here we intend to reveal our progress with both the polarizations. We make use of the following ansatz of the invariant corresponding to the Hamiltonian in eq.(\ref{2.9}). 
\begin{equation}\label{a.6}
\begin{split}
\hat{I}(t)&=\alpha_1(t)\hat{p}_1^2+\alpha_2(t)\hat{p}_2^2+\beta_1(t)\hat{x}_1^2+\alpha_2(t)\hat{x}_2^2+\delta_1(t)(\hat{x}_1\hat{p}_1+\hat{p}_1\hat{x}_1)+\delta_2(t)(\hat{x}_2\hat{p}_2+\hat{p}_2\hat{x}_2)+\lambda_1(t)\hat{x}_1\hat{p}_2+\lambda_2(t)\hat{p}_1\hat{x}_2\\&+\lambda_3(t)\hat{p}_1\hat{p}_2+\lambda_4(t)\hat{x}_1\hat{x}_2
\end{split}
\end{equation}
where $\alpha_1,\alpha_2,\beta_1,\beta_2,\delta_1,\delta_2,\lambda_1,\lambda_2,\lambda_3$, and $\lambda_4$ are all unknown time dependent parameters. 
Computing $i\hbar\frac{\partial \hat{I}(t)}{\partial t}$ and $[\hat{I},\hat{H}]$, we obtain a set of ordinary differential equations given by
\begin{align}
&\dot{\alpha}_1-4\alpha_1d+2\delta_1a-\lambda_3f=0~,\label{a.7.1}\\
&\dot{\alpha}_2+4\alpha_2d+2\delta_2a-\lambda_3f=0~,\label{a.7.2}\\
&\dot{\beta}_1+4\beta_1d-2\delta_1b+\lambda_4f=0~,\label{a.7.3}\\
&\dot{\beta}_2-4\beta_2d-2\delta_2b+\lambda_4f=0~,\label{a.7.4}\\
&\dot{\delta}_1-\alpha_1b+\beta_1 a-\lambda_1f=0~,\label{a.7.5}\\
&\dot{\delta}_1-\alpha_1b+\beta_1 a+\lambda_2f=0~,\label{a.7.6}\\
&\dot{\delta}_2-\alpha_2b+\beta_2 a+\lambda_1f=0~,\label{a.7.7}\\
&\dot{\delta}_2-\alpha_2b+\beta_2 a-\lambda_2f=0~,\label{a.7.8}\\
&\dot{\lambda}_1-2\delta_1f+2\delta_2f+4\lambda_1d-\lambda_3b+\lambda_4a=0~,\label{a.7.9}\\
&\dot{\lambda}_2+2\delta_1f-2\delta_2f-4\lambda_2d-\lambda_3b+\lambda_4a=0~,\label{a.7.10}\\
&\dot{\lambda}_3-2\alpha_1f-2\alpha_2f+(\lambda_1+\lambda_2)a=0~,\label{a.7.11}\\
&\dot{\lambda}_4+2\beta_1f+2\beta_2f-(\lambda_1+\lambda_2)b=0~.\label{a.7.12} 
\end{align}
Comparing eq.(\ref{a.7.5}) with eq.(\ref{a.7.6}) (or eq.(\ref{a.7.7}) with eq.(\ref{a.7.8})), one can easily find out that $\lambda_1=-\lambda_2=\lambda_0$. Substituting $\lambda_1=-\lambda_2=\lambda_0$ in eq.(s)(\ref{a.7.9},\ref{a.7.10}), we obtain one differential equation and a constraint equation given by
\begin{align}
&\dot{\lambda}_0-2\delta_1f+2\delta_2f=0~,\label{a.7.13}\\
&4\lambda_0d-\lambda_3b+\lambda_4a=0\label{a.7.14}~.
\end{align} 
Next we use the parametrization $\alpha_1(t)=\kappa_1\rho_1^2(t)~,~\alpha_2(t)=\kappa_2\rho_2^2(t)$, where $\kappa_1$ and $\kappa_2$ are two time independent unknown constants. Then, in terms of $\rho_1$ and $\rho_2$, we can obtain the other time dependent coefficients up to $\mathcal{O}(h)$ as
\begin{align}
\beta_1(t)&=\frac{\kappa_1}{a^2}\left(\dot{\rho}_1^2+\rho_1\ddot{\rho}_1-2\rho_1^2\dot{d}-4\rho_1\dot{\rho}_1d+ab\rho_1^2\right)~,\label{a.9.1}\\
\beta_2(t)&=\frac{\kappa_2}{a^2}\left(\dot{\rho}_2^2+\rho_2\ddot{\rho}_2+2\rho_2^2\dot{d}+4\rho_2\dot{\rho}_2d+ab\rho_2^2\right)~,\label{a.9.2}\\
\delta_1(t)&=-\frac{\kappa_1}{a}(\rho_1\dot{\rho_1}-2\rho_1^2d)~,\label{a.9.3}\\
\delta_2(t)&=-\frac{\kappa_2}{a}(\rho_2\dot{\rho_2}+2\rho_2^2d)~,\label{a.9.4}\\
\lambda_0(t)&=\frac{2}{a}\int_0^tdt'f(t')(\kappa_2\rho_2(t')\dot{\rho}_2(t')-\kappa_1\rho_1(t')\dot{\rho}_1(t'))~,\label{a.9.5}\\
\lambda_3(t)&=2\int_0^t dt'f(t')(\kappa_1\rho_1^2(t')+\kappa_2\rho_2^2(t'))~,\label{a.9.6}\\
\lambda_4(t)&=\frac{2b}{a}\int_0^t dt'f(t')(\kappa_1\rho_1^2(t')+\kappa_2\rho_2^2(t'))~.\label{a.9.7}
\end{align}
These equations are difficult to solve exactly and hence we go back to the analysis in section \ref{S4}.

\end{document}